\documentclass[aps,prb,twocolumn,longbibliography,showkeys,showpacs,amssymb]{revtex4-2}
\usepackage{latexsym}
\usepackage{amsfonts}
\usepackage{amsmath}
\usepackage{color}
\usepackage{graphicx}
\usepackage{mathptmx}      
\usepackage{subfigure}
\usepackage{listings}
\usepackage{grffile}
\usepackage{bm}
\definecolor{amethyst}{rgb}{0.6, 0.4, 0.8}
\definecolor{DarkGreen}{rgb}{0.00,0.39,0.00}

\newcommand\bB{\mathbf{B}}
\newcommand\bS{\mathbf{S}}

\usepackage{hyperref}
\hypersetup{
    colorlinks   = true, 
    urlcolor     = blue, 
    linkcolor    = blue, 
    citecolor    = blue 
}
\usepackage{xr}
\begin{document}
\title{Sustaining Rabi oscillations by using a phase-tunable image drive}

\author{H. De Raedt}
\email{deraedthans@gmail.com}
\thanks{Corresponding author}
\affiliation{Institute for Advanced Simulation, J\"ulich Supercomputing Centre,
Forschungszentrum J\"ulich, D-52425 J\"ulich, Germany\\}
\affiliation{Zernike Institute for Advanced Materials,
University of Groningen, Nijenborgh 4, NL-9747AG Groningen, The Netherlands}
\author{S. Miyashita}
\affiliation{
Department of Physics, Graduate School of Science,\\
University of Tokyo, Bunkyo-ku, Tokyo 113-0033, Japan}
\author{Kristel Michielsen}
\affiliation{Institute for Advanced Simulation, J\"ulich Supercomputing Centre,\\
Forschungszentrum J\"ulich, 52425 J\"ulich, Germany}
\affiliation{RWTH Aachen University, 52056 Aachen, Germany}

\author{H. Vezin}
\affiliation{
CNRS, Universit\'e de Lille, {\it LASIRE (UMR 8516)}, \\
Laboratoire de Spectroscopie pour les Interactions, la R\'eactivit\'e et l'Environnement,
F-59655 Villeneuve d'Ascq, France}

\author{S. Bertaina}
\affiliation{
CNRS, Aix-Marseille Universit\'e,
{\it IM2NP (UMR 7334)},
Institut Mat\'eriaux Micro\'electronique et Nanosciences de Provence,  F-13397 Marseille, France}

\author{I. Chiorescu}
\affiliation{
Department of Physics, The National High Magnetic Field Laboratory,
Florida State University, Tallahassee, Florida 32310, USA}

\date{\today}

\begin{abstract}
{\color{black}
Recent electron spin resonance experiments on CaWO$_4$:Gd$^{3+}$ and on other magnetic impurities have demonstrated
that sustained Rabi oscillations can be created by driving a magnetic moment with a microwave field frequency slightly larger than the Larmor frequency and tuned to the Floquet resonance
together with another microwave field (image drive) with a frequency
smaller than the Larmor frequency.
These observations are confirmed by the new experimental results reported in this paper. To study the interplay between the microwave drives and the decay due to decoherence and dissipation, we investigate several mechanisms of the latter by using a combination of numerical and analytical techniques.
Concretely, we study three different sources of decoherence and dissipation.
The first microscopic model describes a magnetic moment in external magnetic fields, interacting with a bath of two-level systems acting as a source of decoherence and dissipation.
The second model describes a collection of the identical, interacting  magnetics moments, all subject to the same magnetic fields.
In this case, the many-body interactions causes a decay of the Rabi oscillations.
In addition, we also study the effect of the inhomogeneity of the microwave radiation on the decay of the Rabi oscillations.
Our simulation results show that the dynamics of a magnetic moment
subject to the two microwave fields with different frequencies
and in contact with an environment is highly nontrivial.
We show that under appropriate conditions, and in particular at the Floquet resonance, the magnetization exhibits sustained  Rabi oscillations, in some cases with additional beatings.
Although these two microscopic models separately describe the experimental data well, a simulation study that simultaneously accounts for  both types of interactions is currently prohibitively costly.
To gain further insight into the microscopic dynamics of these two different models,
we study the time dependence of the bath and system energy and of
the correlations of the spins, data that is not readily accessible experimentally.
}
\end{abstract}


\maketitle
\clearpage

\section{Introduction}\label{section0}

The implementation of quantum processors requires precise control of quantum
states exhibiting long decoherence times $T_2$ in order to perform quantum
algorithms or construct quantum memories. However, interactions with
environmental degrees of freedom limit the coherence time and the qubit
operability. In the case of spin qubits in solid state materials, examples of
decoherence sources are spin-spin interactions, spin diffusion, fluctuations in
the magnetic field or charge/electric noise
\cite{Chirolli2008,Yoneda2018a}. These interactions lead to a pure
dephasing time $T_\phi$ of the qubit much shorter than its energy relaxation
time $T_1$.

A number of techniques have been developed recently aiming to replace the pure
dephasing with an actively controlled decoherence time by means of Electron Spin
Resonance (ESR) pulses such as Dynamical Decoupling (DD)
\cite{Viola1998,Viola1999,Viola1999a,Uhrig2007,Khodjasteh2005} and Concatenated
Dynamical Decoupling (CDD) \cite{Cai2012,Cohen2017}. The former uses a sequence
of spin flips to improve spin coherence via a spin-echo technique, while the
latter is using additional layer(s) of DD to solve for the imperfections and
fluctuations of the first layer. The methods make use of strong $\pi$-pulses
which complicates the operations due to gate imperfections; similar issues are
noted in the case of a proposal using strong continuous waves
\cite{Facchi2004,Fanchini2007}. In general, DD techniques applied to electronic spin qubits have been used
on nitrogen vacancy (NV) centers and with relative success
\cite{Cai2012,Farfurnik2017,Teissier2017,Rohr2014}.

However, other spin qubit implementations with different anisotropic/isotropic
properties \cite{Bertaina2009a,Zeisner2019,Soriano2022}, spin levels \cite{Bertaina2009},
electro-nuclear configurations \cite{Bertaina2017}, and host materials need to
be studied for their potential use in quantum computing and quantum memories.
Rare earth ions such as CaWO$_4$:Er$^{3+}$ \cite{LeDantec2021},
Y$_2$SiO$_5$:Er$^{3+}$ \cite{Welinski2019} and Y$_2$SiO$_5$:Yb$^{3+}$
\cite{Lim2018} or $^{28}$Si:Bi \cite{Bienfait2016} reach long relaxation times
as well as significant coherence times. Therefore, a universal method is needed,
applicable to any type of qubit (spin or superconducting qubit for instance),
using weak pulses which do not alter a quantum algorithm. In
Ref.~\cite{BERT20} we demonstrate a protocol based on Floquet resonances
\cite{Russomanno2017} using two microwave drives acting in parallel on the
quantum state of a qubit and able to increase the coherence time under drive  up to the
relaxation time within the experimental conditions of
Ref.~\cite{BERT20}.  The main drive is induces imperfect Rabi oscillations while
a circularly polarized second drive, of about two orders of magnitude smaller in
amplitude, has the role of sustaining the oscillations. Rather than decoupling
the qubit from the bath using a strong excitation, we use very weak pulses and
alter the dynamics of the entire system. In practice, the time interval between two regular pulsed gates could contain
an integer number of Rabi periods protected by our protocol and therefore the quantum information would be preserved from one gate to the next.  This method was demonstrated
in Ref.~\cite{BERT20} on any initial state of spin systems with
different anisotropies, spin sizes and spin-orbit couplings. The protocol is universal and it should be applicable to other qubit implementations such as nuclear spins or superconducting qubits. Similar protocols
have been recently implemented for the case of SiC divacancy defects
\cite{Miao2020} and NV centers \cite{Wang2021,Wang2020}.

{\color{black}
In this paper, the experimental protocol described in Ref.~\cite{BERT20} is
studied analytically and by computer simulation.
The aim of this paper is to scrutinize the microscopic physical mechanisms that may cause the
sustained, long-time Rabi oscillations to appear and to give a consistent
microscopic description of the experimental data displayed in Fig.~\ref{FIGexp}.

We consider three different models for the source of decoherence..
The first model describes a spin qubit, that is a magnetic moment,
subject to external magnetic fields and interacting with a bath of two-level systems.
These two-level systems are not directly affected by the external magnetic fields.
This kind of bath mimics an environment consisting of two-level systems
representing e.g., nuclear spins, electronic spins of different species, defects, etc.,

The second model describes an ensemble of identical magnetic moments, all driven by the same
microwave fields and interacting with each other via all-to-all, dipolar-like weak couplings.
The effect of these couplings on any of the magnetic moments is to cause decoherence.
Thus, from the viewpoint of an arbitrarily chosen magnetic moment, all others may
be regarded as representing a kind of ``bath''.
The third model extends the second one by taking the inhomogeneity of the microwave fields into account.

In the first model, the spins of the bath act as bath degree of freedom and we refer to it as ``bath-I''.
In the second model, the interactions among the spins cause decoherence due to many-body effects. In this case, the ensemble itself play a role of the bath and we refer to it as `bath-II''. In the third model, the inhomogeneity is an additional source of decoherence.
We mainly study the first and second models and study the third one in combination with the second model.
}

The paper is organized as follows. In Section II, we present
and discuss new electron spin resonance results for CaWO$_4$:Gd$^{3+}$.
Section III gives an overview of the three different models that are at the basis of our simulation work.
In Section IV, we specify the bath-I model in detail and present the simulation results.
Section V introduces the Hamiltonian of the bath-II model and discusses
the simulation results for the case without and with the inhomogeneity of the microwave fields.
In Section VII, we present are conclusions.
Technical details and additional results can be found in
the Supplemental Material \footnote[1]{See Supplemental Material at [URL] for calculation details: Transformation to the rotating frame, Schrodinger dynamics of the single-spin system,Floquet theory, quantum master equation approach, TDSE simulations, spin-spin correlations,}.


\section{Electron spin resonance experiments}\label{EXP}

Experiments probing the magnetization dynamics under external driving fields
such as the ones performed in Ref.~\cite{BERT20}, pose interesting
theoretical problems.
In these electron spin resonance (ESR) experiments, one measures a signal that
is proportional to the expectation value of the $z$-component of the magnetization
where the $z$ direction is defined by the direction of the static magnetic field $\bB_0$.
In symbols:
\begin{eqnarray}
\mathrm{Signal}&\propto&
\sum_{n=1}^{N}\langle S^z_n(t) \rangle
,
\label{ESR0}
\end{eqnarray}
where $N$ is the number of magnetic moments in the sample.
The presence of a high frequency resonant microwave field perpendicular to the
static field induces Rabi oscillations. A characteristic feature of Rabi
oscillations is that their frequency changes linearly with the applied microwave power.
These oscillations decay due to various dissipation effects. Recently, it was
demonstrated that sustained Rabi oscillations can be realized by using an image
drive of the microwave field which itself is slightly detuned from
resonance~\cite{BERT20}.

In the ESR experiments under scrutiny~\cite{BERT20},
Rabi oscillations are induced by applying an electromagnetic
drive pulse of strength $h_d$ and frequency $f=f_0+\Delta$ where $f_0$ is the
Larmor frequency and $\Delta$ is a (small) detuning~\cite{BERT20}.
Simultaneously with the drive pulse, an image pulse of strength $h_i$ and of
frequency $f=f_0-\Delta$, coherent with the drive pulse, is
applied~\cite{BERT20}. The power of the image drive is much less than that of
the drive pulse itself, i.e. $h_i\ll h_d$. The phase difference $\phi$ between
the drive and image pulse can be used to manipulate the Rabi
oscillations~\cite{BERT20}. Under optimal image drive conditions, experiments
show sustained Rabi oscillations up to the maximum amplifier gate length of
$15\,\mu\hbox{s}$~\cite{BERT20}.

\begin{figure}[!htp]
\centering
\includegraphics[width=\hsize]{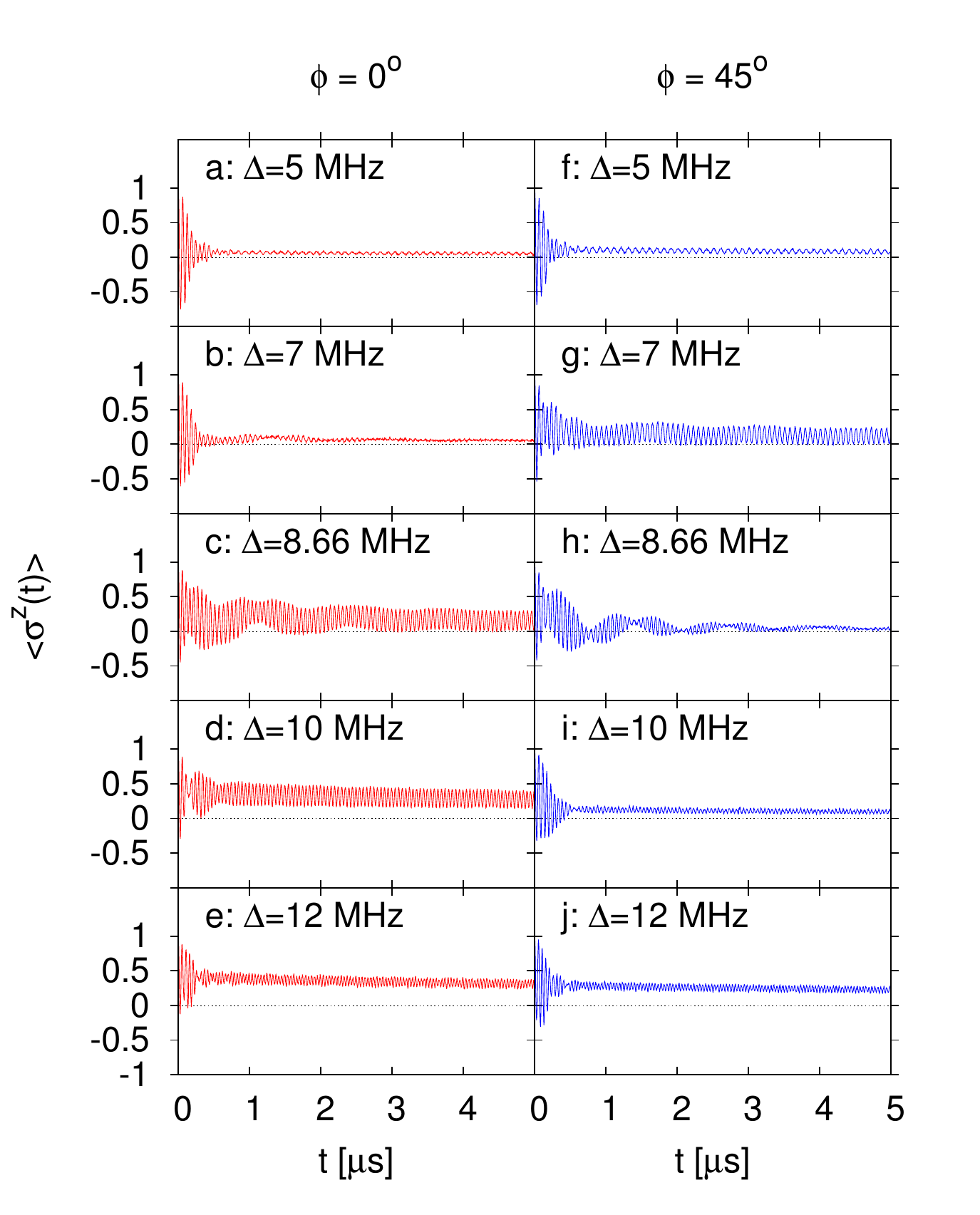}  
\caption{(color online)
Experimental data of the Rabi oscillations in CaWO$_4$:Gd$^{3+}$ at $T=40\,\mathrm{K}$
for different values of the detuning $\Delta$, phase differences $\phi=0$ and $\phi=45^\circ$,
and fixed amplitudes $h_d=15\,\hbox{MHz}$ and $h_i=0.12h_d=1.8\,\hbox{MHz}$, respectively.
\label{FIGexp}
}
\end{figure}

Beside data previously reported in Ref.~\cite{BERT20}, we present new experiments in Fig.~\ref{FIGexp}.
These data have been obtained for Gd$^{3+}$ moments 
in a CaWO$_4$ host lattice.
Although these moments have spin $S=7/2$, the orientation of the static
magnetic field and the frequency and power of the microwave pulses are chosen to
select transitions between one pair of Zeeman levels only~\cite{BERT20}.
Therefore, the spin system may be viewed as an effective two-level system performing Rabi oscillations.
In this paper, we take the CaWO$_4$:Gd$^{3+}$ system for our case study and use
the results presented in Fig.~\ref{FIGexp} as reference for comparison with the model calculations.
In this paper, we use Fig.~\ref{FIGexp} as a reference for comparing with our simulation results for the three different models.

The main conclusion drawn on the basis of the experimental data~\cite{BERT20}
and new experimental data, some of which are shown in Fig.~\ref{FIGexp}
may be summarized as follows:
\begin{enumerate}
\item
{\color{black}
Regarding the effect of the image pulse on the decay of the Rabi oscillations,
the experimental findings are fairly generic, meaning that the main features do not significantly depend
on the kind of impurity that consitutes the qubit, the presence or absence of nuclear spins, etc.
See Ref.~\onlinecite{BERT20} for more details.}
\item
{\color{black}
In the absence of the image pulse, the amplitudes of the Rabi oscillations vanish rapidly,
on a time scale of $1\mu\mathrm{s}$~\cite{BERT20}.
}
\item
For a detuning $\Delta=5\,\mathrm{MHz}$,
the Rabi oscillations decay rapidly, on a time scale of $1\mu\mathrm{s}$
see Figs.~\ref{FIGexp}(a,f).
\item
For a detuning $\Delta=7\,\mathrm{MHz}$,
we observe either Rabi oscillations that decay rapidly, on a time scale of $1\mu\mathrm{s}$,
if $\phi=0$ (see Fig.~\ref{FIGexp}(b))
or sustained Rabi oscillations
if $\phi=45^\circ$ (see Fig.~\ref{FIGexp}(g))
\item
For a detuning equal to the detuning
$\Delta=8.66\,\hbox{MHz}$ that corresponds to the first Floquet resonance frequency (see below),
we observe sustained Rabi oscillations with some
signatures of beating, Figs.~\ref{FIGexp}(c,h),
depending on the value of the phase $\phi$ between the drive and image pulse, in qualitative agreement
with earlier experiments on MnO:Mg$^{2+}$~\cite{BERT20}.
\item
Increasing the detuning further to $\Delta=10,12\,\hbox{MHz}$,
see Figs.~\ref{FIGexp}(d,e,i,j), we observe sustained Rabi oscillations with
small amplitudes.
\item
Figures~\ref{FIGexp}(c,d,g) show that the appearance of sustained Rabi oscillations
depends on both $\Delta$ and $\phi$.
Therefore this appearance cannot be solely attributed
to the existence of a Floquet resonance, the frequency of which depends
on $\Delta$ and $h_d$ and is independent of $\phi$ (see below).
\end{enumerate}

In summary, from the experimental results shown in Fig.~\ref{FIGexp}, we conclude that there are two striking features.
Depending on the value of the detuning further to $\Delta$
(for constant drive and image drive amplitudes $h_d$ and $h_i$, respectively).
sustained long-time Rabi oscillations appear. This is the main feature.
In contrast to the usual decaying Rabi oscillations observed in the absence of the detuning ($\Delta=0$),
in some situations the Rabi oscillations may persist for a long time.
The second feature is that these sustained oscillations may exhibit
additional structure depending on the detuning $\Delta$ and the phase shift $\phi$.

{\color{black}
\section{Overview of the microscopic models}\label{section0a}
}

The magnetic moments being probed in the ESR experiments are distributed
in the host lattice in a highly diluted manner~\cite{BERT20}.
Thus, the most basic model is that of non-interacting spins subject
to a time-dependent magnetic field, see section~\ref{section1z}.
In this paper, we study a model in which the contributions to the magnetic field are (1) a strong static magnetic field,
(2) a microwave drive field and (3) the novel aspect, the microwave image field~\cite{BERT20}.
The pure quantum dynamics of this model is straightforwardly studied by solving the time-dependent Schr\"odinger
equation for a single spin, see section~\ref{section1z}.
To account for the explicit time-dependence of the system + bath Hamiltonian,
we use a second-order decomposition formula for ordered matrix exponentials~\cite{SUZU93,RAED06} to solve the TDSE.

Obviously, there is no decoherence or dissipation in this most basic model.
In the presence of the image drive, the magnetization dynamics exhibits modulated Rabi oscillations.
The amplitude of the modulation depends on the detuning $\Delta$ and the phase shift $\phi$.

In the present paper, we study the decay of the Rabi oscillations by solving for the pure quantum dynamics of models with many spins in which decoherence and dissipation appear automatically.
Decoherence and dissipation effects may also be studied through quantum master equations~\cite{BREU02}.
This approach makes several implicit assumptions and often involves many parameters~\cite{BREU02}.
We found it difficult to find a set of parameters which reproduce,
even qualitatively, the dependence on the detuning $\Delta$ and the phase shift $\phi$ observed in experiments.
In section~\ref{A-SUP.DECO} of the Supplemental Material~\cite{Note1},
it is shown that, even when the inhomogeneity of the microwave field is included, this method does not describe all the qualitative features of the experimental data shown in Fig.~\ref{FIGexp}.
Therefore, in this paper we focus on microscopic many-body models
and explore the conditions under which these models
reproduce the systematic trends of the features observed experimentally.

As alluded to earlier, in real samples, the microscopic mechanisms accounted for
separately in the present paper are most likely simultaneously active.
Unfortunately, solving the time-dependent Schr\"odinger equation for
say 28 mutually interacting spins interacting with 28 two-level systems
requires computational resources which at present are prohibitive.
Performing such simulations may be a challenging project for the future.

In the next three sections, we scrutinize theoretical models by visually comparing the outcomes of numerical
simulations with the experimental results shown in Fig.~\ref{FIGexp}.

\medskip
\section{Single-spin system}\label{section1z}

This section introduces the model for the magnetic moment of a single Gd$^{3+}$ ion subject to external magnetic fields.
To good approximation (for our purposes), this model describes a two-level system driven by a periodic field.
Although too simple to serve as a model for the system studied experimentally, its dynamics is already complicated.
For instance, because of the presence of a driving fields $h_d$ and $h_i$, the model can exhibit resonances and beating oscillations.
Of central importance for the present study is the existence of
what will be referred to as the first Floquet resonance (see below).
For pedagogical purposes, this section presents a few results of the dynamics of
the single-spin model in time-dependent magnetic fields.

The system (S), contains one magnetic moment in an external time-dependent magnetic
field and is defined by the Hamiltonian
\begin{eqnarray}
H_\mathrm{S}&=&
-\omega_0 S^z - 2\omega_d S^x\sin[2\pi(f_0+\Delta) t + \phi]
\nonumber \\
&&- 2\omega_i S^x\sin[2\pi(f_0-\Delta) t - \phi]
,
\label{SCB1}
\end{eqnarray}
where $\mathbf{S}=(S^x,S^y,S^z)$ denotes the three components of magnetic moment.
We use units such that $\hbar=1$ and that $f_0=\omega_0/2\pi$, $\Delta$,
$h_d=\omega_d/2\pi$, and $h_i=\omega_i/2\pi$ are frequencies expressed in MHz.
The first term in Eq.~(\ref{SCB1}) describes the Larmor rotation (frequency $f_0$) of the spin due the static magnetic field $\bB_0$.
The second and third term in Eq.~(\ref{SCB1}) are called the microwave drive and image field, respectively.

In ESR experiments, $\omega_0$ typically is several orders of magnitude larger than $\omega_d$ and $\omega_i$.
Instead of adopting the standard resonance condition,
we choose $\omega=\omega_0+2\pi\Delta=2\pi(f_0+\Delta)$, see Ref.~\cite{BERT20},
and use $U=\exp\left(i\omega t S^z\right)$ to find
(for details on the calculation, see section~\ref{A-SUP.TRF} of the Supplemental Material~\cite{Note1})
\begin{eqnarray}
H_\mathrm{S,RF}(t)&\equiv&-ia^\dagger\frac{\partial}{\partial t}U+U^\dagger H_\mathrm{S} U
\nonumber \\
&=&2\pi\big\{
\Delta S^z
- h_d \left(S^x \sin\phi + S^y\cos\phi\right)
\nonumber \\
&&+ h_i \left[S^x \sin(4\pi t\Delta+\phi) - S^y \cos(4\pi t \Delta+ \phi) \right]
\big\}
\nonumber \\
&=&-\bB(t)\dot\bS
\;,
\label{TRF4}
\end{eqnarray}
where we have omitted terms that oscillate with the (very large) frequency $2f_0$
and introduced the time-dependent magnetic field
\begin{eqnarray}
\bB(t)=2\pi
\left(\begin{array}{c}
h_d\sin\phi-h_i\sin(4\pi t\Delta+\phi)\\
h_d\cos\phi+h_i\cos(4\pi t\Delta+\phi)\\
\Delta\\
\end{array}
\right)
\;.
\label{Be}
\end{eqnarray}

If $h_i=0$, $\bB(t)$ is constant in time and it follows immediately that the spin $\bS$
will perform oscillations with the Rabi frequency $F_{\mathrm{R}}=(\Delta^2+h_d^2)^{1/2}$.
Furthermore, if $\Delta=0$, the magnetization $\langle S^z(t)\rangle$ displays the usual Rabi oscillations,
that is the spin performs rotations about the vector $(h_d\sin\phi,h_d\cos\phi,0)^{\mathbf{T}}$.

In section~\ref{A-SUP.TRF} of the Supplemental Material~\cite{Note1}, we show that if $2\Delta=F_R$
(or equivalently $\Delta=h_d/\sqrt{3}$)
and $h_i\not=0$, the spin performs a second kind of rotation with a frequency $F_R^{(2)}=3h_i/4$.
In other words, the condition $2\Delta=F_R$ defines a special point in the
parameter space of the model defined by Eq.~(\ref{TRF4}).

As $\bB(t)=\bB(t+1/2\Delta)$, the Hamiltonian Eq.~(\ref{TRF4}) is a periodic function of time.
Some insight into the dynamics of the periodically driven system described by Eq.~(\ref{TRF4})
can be obtained by resorting to Floquet theory~\cite{FLOQ83,GRIF98}.
It is easy to show (for details see section~\ref{A-SUP.FLOQ} of the Supplemental Material~\cite{Note1})
that in the perturbation regime $|h_i|\ll h_d$, Floquet theory predicts the existence of
resonances if the parameters $\Delta$ and $h_d$ satisfy the condition
\begin{eqnarray}
\Delta=\frac{h_d}{\sqrt{4n^2-1}}\quad,\quad n=1,2,\ldots
\;.
\label{FLOQ1}
\end{eqnarray}
For $n=1$, Eq.~(\ref{FLOQ1}) yields $\Delta=h_d/\sqrt{3}$ and therefore
we refer to $2\Delta=F_R$ as the condition for the first Floquet resonance.  {\color{black} It is essential to note that the Floquet resonance opens up a gap or level repulsion for the quasi-energies in resonance which can be seen as a decoherence dynamical sweet spot: the levels are flat and thus insensitive to noise in $h_d$ or $\Delta$. However, our work doesn’t stop at such qualitative interpretation. Our numerical simulations discover the microscopic, physical mechanisms developing inside the bath (resonant or non-resonant to the drive).}

Experimentally reasonable values for the CaWO$_4$:Gd$^{+3}$ system are
$h_d=15\,\hbox{MHz}$ and $h_i=0.12h_d=1.8\,\hbox{MHz}$~\cite{BERT20}
and the condition for the first Floquet resonance reads
$\Delta=h_d/\sqrt{3}=8.66\,\hbox{MHz}$ with $F_{\mathrm{R}}=2\Delta=17.33\,\hbox{MHz}$~\cite{BERT20}.

\begin{figure}[!htp]
\centering
\includegraphics[width=\hsize]{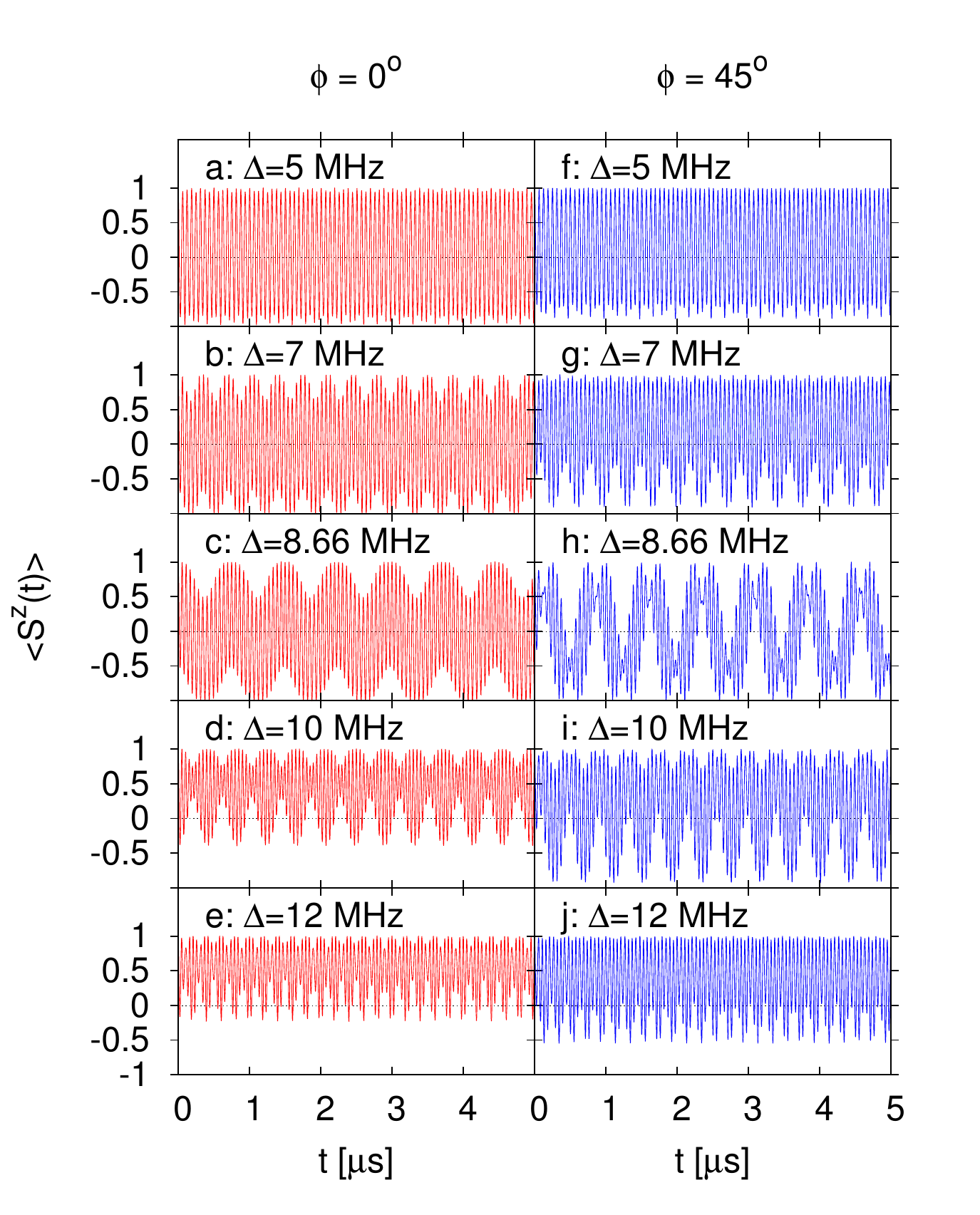}
\caption{(color online)
Simulation results obtained by solving the TDSE 
for the Hamiltonian Eq.~(\ref{TRF4}), describing
a single spin subject to the time-dependent magnetic field
$\bB(t)=2\pi(h_d\sin\phi-h_i\sin(4\pi t\Delta+\phi), h_d\cos\phi+h_i\cos(4\pi t\Delta+\phi),\Delta)^{\mathrm{T}}$,
with the amplitudes $h_d=15\,\hbox{MHz}$ and $h_i=0.12 h_d=1.8\,\hbox{MHz}$.
For additional information, see sections~\ref{A-SUP.TRF}--\ref{A-SUP.SCHROD}
of the Supplemental Material~\cite{Note1}.
\label{FLOGS1a}
}
\end{figure}

The dynamics of the single-spin system Eq.~(\ref{TRF4}) is most readily studied by solving the TDSE
with Hamiltonian Eq.~(\ref{TRF4}) numerically.
In Fig.~\ref{FLOGS1a}, we present simulation results for the Rabi oscillations
for different values of the detuning $\Delta$ and two values of the phase shift $\phi$
(see section~\ref{A-SUP.SCHROD} of the Supplemental Material~\cite{Note1} for
the corresponding pictures in the frequency domain).

As expected, the effect of the image drive is most pronounced if
the parameters match the condition for the first Floquet resonance.
Then, the magnetization exhibits considerable beating, a manifestation of the presence of the second
Rabi frequency $F_R^{(2)}=3h_i/4\ =1.35\,\hbox{MHz}$.

The Fourier transform of the data depicted in Fig.~\ref{FLOGS1a}(c)  
(see Fig.~\ref{A-FTFLOGS1a}(c) of the Supplemental Material~\cite{Note1})
shows a peak at the Rabi frequency $F_R=17.33\,\hbox{MHz}$ and
a lower/upper sideband at $F_R\mp\delta f$ with $\delta f=1.34\,\hbox{MHz}$.
The spectral weight of these sidebands is about 8\% of the main signal.
In addition, there is a weak signal (spectral weight $\approx0.8\,\%$)
at zero frequency with side bands (spectral weight $\approx1\,\%$)
at $\delta f=1.34\,\hbox{MHz}\;\approx F_R^{(2)}=1.35\,\hbox{MHz}$.
From Table~\ref{A-FLOQtab1}, third row, if follows that
$F_R=\kappa_2+\kappa_1 = 17.32 \,\hbox{MHz}$ and $\delta f\approx \kappa_2-\kappa_1 = 1.34 \,\hbox{MHz}$.
The frequency of the modulation in the time dependent signal Fig.~\ref{FLOGS1a}(c)   
is approximately $1.34\,\hbox{MHz}$, the same as to the difference between the Rabi frequency and the frequency the sidebands.
Thus, we can relate the Rabi frequency and the sideband frequencies to the quasi-energies obtained
from Floquet theory and the modulation of the time dependent signal.
Moreover, the sideband frequencies are, to a very good approximation, given by
$F_R^{(2)}=1.35\,\hbox{MHz}$.

From Fig.~\ref{FLOGS1a} and the pictures of the motion of the magnetization on
the Bloch sphere shown in Figs.~\ref{A-SPHEREfig1a} and~\ref{A-SPHEREfig1b},
it is clear that the presence of an image drive can
affect the Rabi oscillations in a nontrivial manner.
The salient features of the magnetization dynamics, that is the Rabi oscillations
and the frequency of the amplitude modulation (if present) of these oscillations
caused by the image drive can be related to quasi-energies obtained from Floquet theory,
see section~\ref{A-SUP.SCHROD} of the Supplemental Material~\cite{Note1}).

In summary, the key point, advanced in Ref.~\cite{BERT20} and illustrated by
Fig.~\ref{FIGexp}, is that the image drive
can be used to significantly reduce the decay time of the Rabi oscillations.

A simple, phenomenological approach to include effects of dissipation and decoherence
is to resort to a description in terms of the Markovian quantum master equation.
In section~\ref{A-SUP.DECO} of the Supplemental Material~\cite{Note1},
it is shown that also this approach, even when the inhomogeneity of the microwave field
is included, does not describe all the qualitative features of the experimental data shown in Fig.~\ref{FIGexp}.


\section{Single-spin system interacting with a bath of two-level systems}\label{section1a}

In this section, we study a microscopic model in which the spin qubit, the magnetic moment
of the Gd$^{+3}$ ion referred to as system (S), interacts with a bath (B) of pseudo-spins.
In this section, the spins other than that of the Gd$^{+3}$ ion are the bath spins.
We analyze the spin dynamics of this many-body $S=1/2$ system by solving the TDSE numerically.
Here,
 $\bS=\bm\sigma/2$ where $\bm\sigma=(\sigma_x,\sigma_y,\sigma_z)=(\sigma_1,\sigma_2,\sigma_3)$
are the Pauli matrices.

The Hamiltonian of the system (S) + bath (B) takes the generic form
\begin{eqnarray}
H(t)&=& H_\mathrm{S}(t) + H_\mathrm{B} + \lambda H_\mathrm{SB}
,
\label{SCB0}
\end{eqnarray}
where $H_\mathrm{B}$ and $H_\mathrm{SB}$ are the bath and system-bath Hamiltonians, respectively.
The overall strength of the system + bath-I interaction is controlled by the parameter $\lambda$.
The Hamiltonian for the system-bath interaction is chosen to be
\begin{eqnarray}
H_\mathrm{SB}&=& -\sum_{n=1}^{N_{\mathrm{B}}} \;\;\sum_{\alpha=x,y,z}J_{n}^\alpha I_{n}^\alpha S^\alpha
,
\label{SCB2}
\end{eqnarray}
where $N_{\mathrm{B}}$ is the number of spins in the bath,
the $J^\alpha_{n}$ are uniform random frequencies in the range $[-J,+J]$ and
$I_{n,k}$ is the $k$-th component if the bath spin $\mathbf{I}_n$.
As the system-bath interaction strength is controlled by $\lambda$, we may set $J=1\,\hbox{MHz}$
without loss of generality.
Note that according to Eq.~(\ref{SCB2}), the spin qubit 
interacts with all the $N_{\mathrm{B}}$ bath spins.
For the bath-I Hamiltonian, we take
\begin{eqnarray}
H_\mathrm{B}&=& -\sum_{n=1}^{N_{\mathrm{B}}}\;\;\sum_{\alpha=x,y,z} K_{n}^\alpha I_{n}^\alpha I_{n+1}^\alpha
,
\label{SCB3}
\end{eqnarray}
where the $K_{n,k}$'s are uniform random frequencies in the range $[-K,K]$.
In our simulation work, we use periodic boundary conditions $I_{n}^\alpha=I_{n+N}^\alpha$.
The Hamiltonian Eq.~(\ref{SCB3}) described a collection of magnetic moments,
located on a ring of lattice sites, and interacting with their nearest neighbors.
Because of the random couplings,
it is unlikely that Eq.~(\ref{SCB3}) is integrable (in the Bethe-ansatz sense)
or has any other special features such as a conserved magnetization etc.

Expressing the motion of the spin qubit 
in the rotating frame has no effect on the bath Hamiltonian Eq.~(\ref{SCB3}) because the transformation to
the rotating frame only affects the spin $\bS$, not the operators describing bath-I.
However, expressing the motion of the spin qubit 
in the rotating frame changes the system-bath Hamiltonian Eq.~(\ref{SCB2}) to
\begin{eqnarray}
H_\mathrm{SB}&=&
-\sum_{n=1}^{N_{\mathrm{B}}} J_{n}^z I_{n}^zS^z
\nonumber \\
&&-\sum_{n=1}^{N_{\mathrm{B}}} J_{n}^x I_{n}^x\left(S^x \cos \omega_0 t+S^y \sin\omega_0 t\right)
\nonumber \\
&&-\sum_{n=1}^{N_{\mathrm{B}}} J_{n}^y I_{n}^y\left(S^y \cos \omega_0 t-S^x \sin\omega_0 t\right)
\;.
\label{SCB2a}
\end{eqnarray}
Discarding the terms that oscillate with the high angular frequency $\omega_0$,
only the coupling between the $z$-component of the system and bath spins survives, yielding
\begin{eqnarray}
H_\mathrm{SB,RF}&=&
-\sum_{n=1}^{N_{\mathrm{B}}} J_{n}^z I_{n}^zS^z
.
\label{SCB2b}
\end{eqnarray}
Therefore, within this approximation, the Hamiltonian of the system (S) + bath (B) in the rotating frame reads
\begin{eqnarray}
H(t)&=& H_\mathrm{S,RF}(t) + H_\mathrm{B} + \lambda H_\mathrm{SB,RF}
.
\label{HA}
\end{eqnarray}
As $H_\mathrm{S,RF}(t)$ and $H_\mathrm{SB,RF}$ do not commute,
the system and bath-I can exchange energy if $\lambda\not=0$.

\begin{figure}[!htp]
\centering
\includegraphics[width=\hsize]{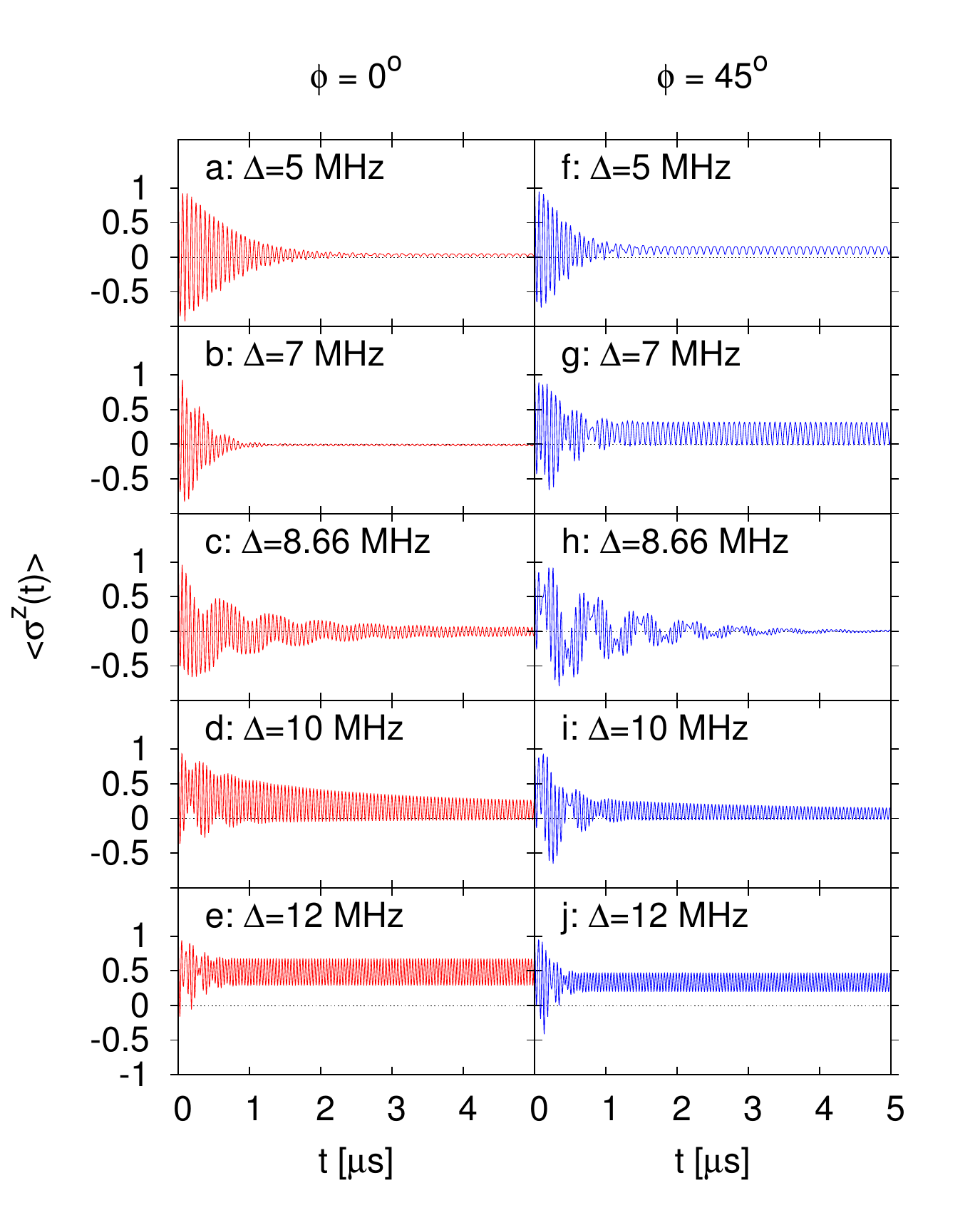}
\caption{(color online)
Simulation results for the magnetization, as obtained by solving the TDSE 
for Hamiltonian Eq.~(\ref{HA}),
describing the system S coupled to bath-I, both subject to a time-dependent magnetic field.
Bath-I contains $N_{\mathrm{B}}=28$ interacting spins,
the spin-bath coupling $\lambda=6$,
$J=1\,\hbox{MHz}$, $K=10\,\hbox{MHz}$, $h_d=15\,\hbox{MHz}$, and $h_i=0.12h_d=1.8\,\hbox{MHz}$.
\label{TDSEfig1a}
}
\end{figure}

{\color{black}
As $J^\alpha_n$ is chosen uniformly random in the interval $[-1\,\hbox{MHz},+1\,\hbox{MHz}]$,
the average of strength of the system-bath interaction is $\lambda\langle |J^\alpha_n|\rangle=\lambda/2 \,\hbox{MHz}$.
On the other hand, the characteristic energy scale of the spin qubit is $2\pi h_d\approx94\,\hbox{MHz}$.
Therefore, for $\lambda<10$, this strength is at least an order of magnitude smaller
than the characteristic energy scale of the spin qubit.
Similarly, for $K=10 \,\hbox{MHz}$ and $\lambda=6$
(see Fig.~\ref{TDSEfig1a})), $\langle |K^\alpha_n|\rangle=5\,\hbox{MHz}$,
of the same order of magnitude as $\lambda\langle |J^\alpha_n|\rangle=3\,\hbox{MHz}$.
}
{\color{black}
In the absence of the image drive, the characteristic decay time
of the Rabi oscillations is $1\mu\mathrm{s}$ or less~\cite{BERT20}.
Performing a simulation with $h_i=0$ and $\Delta=8.66\,\hbox{MHz}$ (data not shown)
we find that the decay time of the Rabi oscillations is about $0.3\mu\mathrm{s}$.
Thus, our choice for the model parameters yields a decay time of the Rabi oscillations
which, in the absence of the image drive, is fairly short and in the ball park
of decay times found experimentally for very different samples~\cite{BERT20}.
}

As explained in more detail in section~\ref{A-SUP.TECH} of the Supplemental Material~\cite{Note1},
to study the Rabi oscillations, the initial state of the whole is taken to be
\begin{eqnarray}
|\Psi(t=0)\rangle&=& |{}\uparrow\rangle \otimes |\Phi\rangle
\;,
\label{IS0}
\end{eqnarray}
where $\Phi$ is a random vector in the $N_{\mathrm{B}}$-dimensional Hilbert space.

Evidently, the energy of the whole system, which is closed, is a conserved quantity.
However, the system S and the bath B are in contact with each other and can exchange energy.
Thus, the system S can show decoherence and relaxation, equilibration, thermalization, all
depending on how the simulation is performed~\cite{ZHAO16,RAED17b}.

Figure~\ref{TDSEfig1a} shows the simulation data obtained by solving the TDSE 
for Hamiltonian Eq.~(\ref{HA}) with a bath of $N_{\mathrm{B}}=28$ spins.
Qualitatively, model Eq.~(\ref{HA}) reproduces all essential features of the experimental data shown in
Fig.~\ref{FIGexp}.

In model Eq.~(\ref{HA}), there are only two adjustable parameters, namely the system-bath interaction
strength $\lambda$ and the scale of the inter-bath interactions $K$.
In view of the fact that solving the TDSE for a system with 29 spins is quite
expensive in terms of computer resources, we did not try to improve the qualitative agreement
with the experimental results by adjusting these parameters.

It should be noted that being exactly at the Floquet resonance frequency
$\Delta=8.66\,\hbox{MHz}$ is not a necessary condition to observe sustained Rabi
oscillations: they also appear in Figs.~\ref{TDSEfig1a}(b,g) (and
Figs.~\ref{FIGexp}(b,g)).

\begin{figure}[!htp]
\centering
\includegraphics[width=0.95\hsize]{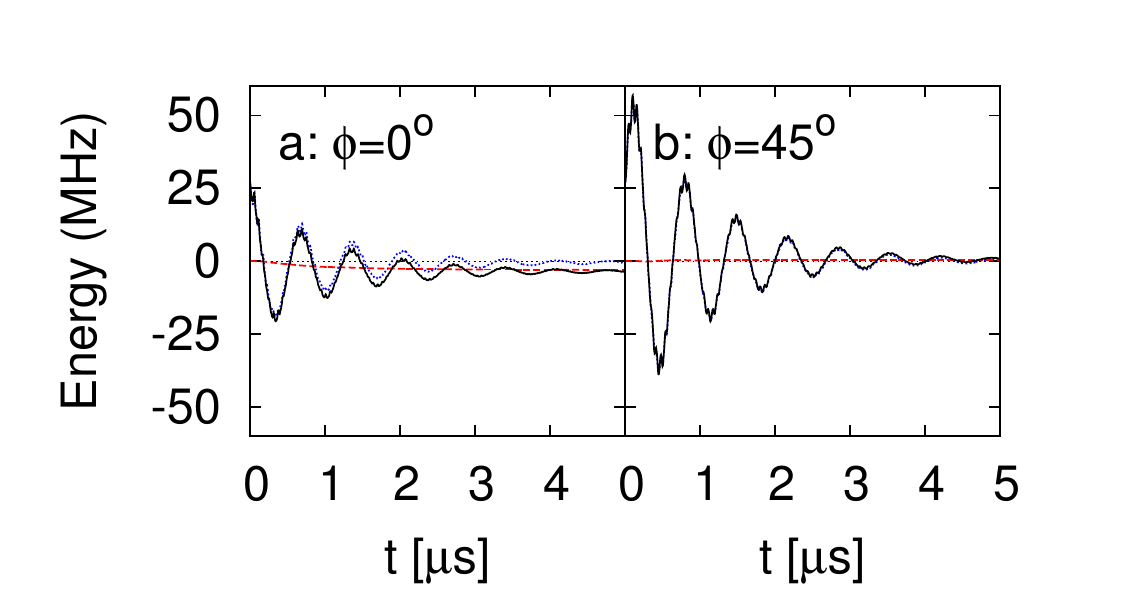}
\caption{(color online)
Simulation results for the energy of the spin bath (dashed red line),
the system (dotted blue line) and total energy (solid black line), as obtained by solving the TDSE 
for Hamiltonian Eq.~(\ref{HA}), describing the single-spin system S coupled to a spin bath, both subject to a time-dependent magnetic field.
The bath consists of $N_{\mathrm{B}}=28$ spins,
the spin-bath coupling $\lambda=6$, $J=1\,\hbox{MHz}$, $K=10\,\hbox{MHz}$, $h_d=15\,\hbox{MHz}$,
$h_i=0.12h_d=1.8\,\hbox{MHz}$, and $\Delta=8.66\,\hbox{MHz}$.
The lines for the system and total energy nearly overlap.
Results for non-resonant values of $\Delta=8.66\,\hbox{MHz}$ are presented in Fig.~\ref{A-SUP.TDSEfig1e}.
\label{TDSEfig1e}
}
\end{figure}

In Fig.~\ref{TDSEfig1e} we present TDSE simulation data of the energy of the system
$E_\mathrm{S}=\langle \Psi(t)|H_\mathrm{S,RF}|\Psi(t)\rangle$, of bath-I
$E_\mathrm{B}=\langle \Psi(t)|H_\mathrm{B,RF}|\Psi(t)\rangle$, and the total energy
$E_\mathrm{S}+E_\mathrm{B}+\lambda E_\mathrm{SB}\langle \Psi(t)|H(t)|\Psi(t)\rangle$.
{\color{black}
From a more general, many-body physics viewpoint, these data show interesting behavior.
For sure, it is not evident that the qubit will evolve to a stationary state
if it is driven by a time dependent field (of strength $h_i$)~\cite{SHIR15,SHIR16}.
Recall that for $\lambda=6$, the qubit-bath interaction is rather weak.
Nevertheless, as Fig.~\ref{TDSEfig1e} shows, in the case under study, it is clear that as time progresses,
both the qubit and bath-I end up in a stationary state, the details of which depend
on the offset $\Delta$ and the phase shift $\phi$.
}

Furthermore, if the value of $\Delta$
matches the resonance condition $\Delta=h_d/\sqrt{3}$, the system and total energy
decay to the stationary state on a much longer time scale than if $\Delta=5,7,10,12\,\hbox{MHz}$
(see Fig.~\ref{A-SUP.TDSEfig1e} of the Supplemental Material~\cite{Note1}).

The period of the large-amplitude oscillations in Fig.~\ref{TDSEfig1e}
is approximately $0.7\,\mu\hbox{s}\approx P^{(2)}_\mathrm{R}=0.74\,\mu\hbox{s}$.
Thus, at resonance ($\Delta=h_d/\sqrt{3}$), the system and total energy exhibit
a synchronized, damped oscillatory behavior with a frequency that is
approximately given by the second Rabi frequency $F^{(2)}_\mathrm{R}=3h_i/4$.

\begin{figure}[!htp]
\centering
\includegraphics[width=\hsize]{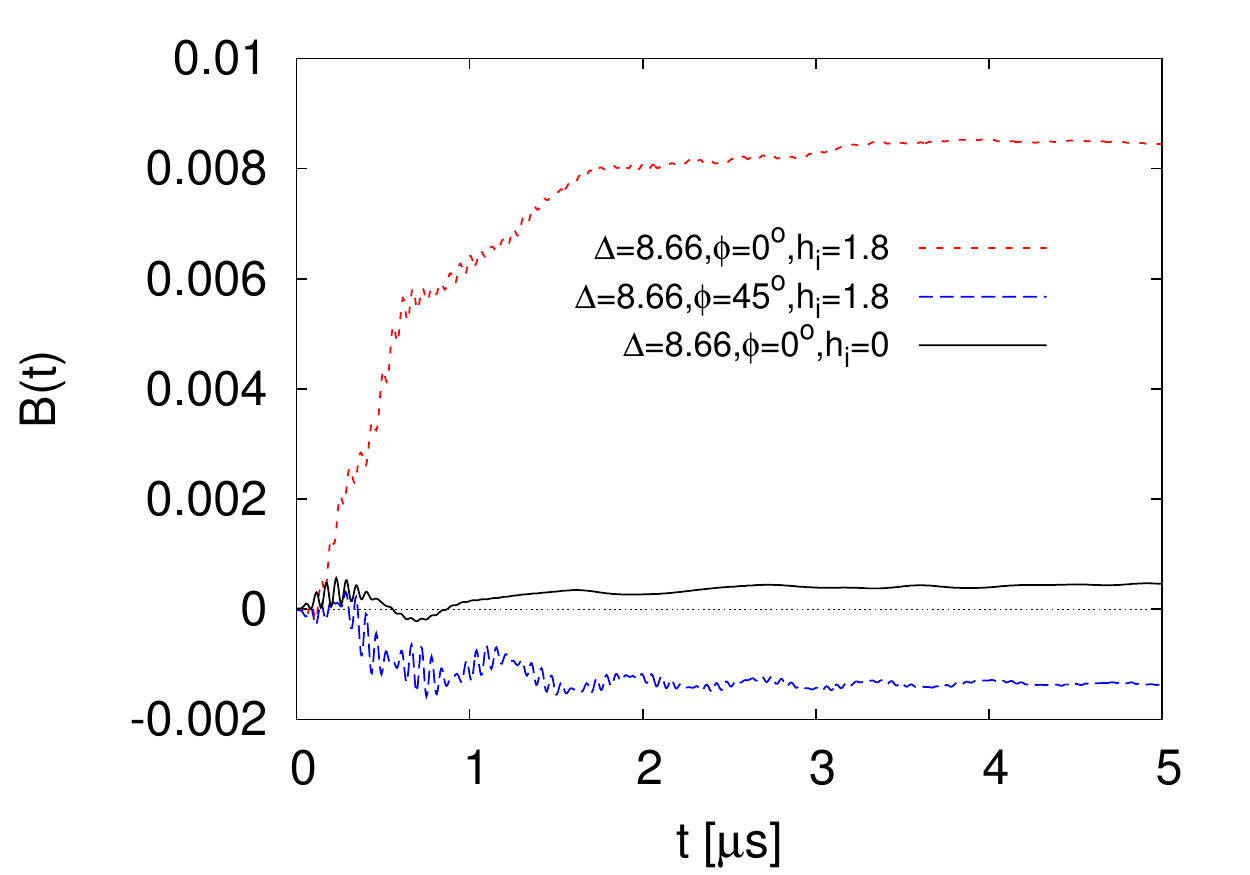}
\caption{(color online)
Simulation results of correlation Eq.~(\ref{SCA1a})
obtained by solving the TDSE 
for Hamiltonian Eq.~(\ref{HA})
describing the single-spin system S coupled to a spin bath, both subject to a time-dependent magnetic field.
The bath consists of $N_{\mathrm{B}}=28$ interacting spins.
\label{model2corr}
}
\end{figure}

In Fig.~\ref{model2corr} we show the correlation function
\begin{equation}
B(t)=\frac{1}{N_{\mathrm{B}}}
\sum_{\alpha=x,y,z}\sum_{n=1}^{N_{\mathrm{B}}}\;
\langle\Psi(t=0)| I_{n}^\alpha(t) I_{n+1}^\alpha(t)|\Psi(t=0) \rangle
\;,
\label{SCA1a}
\end{equation}
that is, the sum of all equal-time, nearest-neighbor correlations of all bath-I pseudo spins
components at the Floquet resonance $\Delta=8.66\,\mathrm{MHz}$.
A first observation is that these correlations are small (relative to the maximum of $B(t)$ which is equal to one).
A second observation, less  clear than in the case $h_i=1.8\,\mathrm{MHz}$ and $\phi=0$,
is that as time proceeds, the correlations seem to saturate.
The transient dynamics includes signatures of the Floquet and Rabi dynamics, because the spin qubit 
is driving bath-I via the $H_{SB}$ interaction.
Also, it's important to note that a saturation of the bath-I internal dynamics
can lead to an increase of the coherence time, as it was observed in the case of
molecular magnets placed in high magnetic fields~\cite{TAKA09}.
The behavior of the bath-I correlation $B(t)$ is
very different from the that of the correlations
in the model of interacting system spins, discussed in the next section.

In summary:
the model Eq.~(\ref{HA}) reproduces the main features (see section~\ref{section0})  of the spin dynamics
observed in the ESR experiments on CaWO$_4$:Gd$^{3+}$.

\begin{figure}[!htp]
\centering
\includegraphics[width=\hsize]{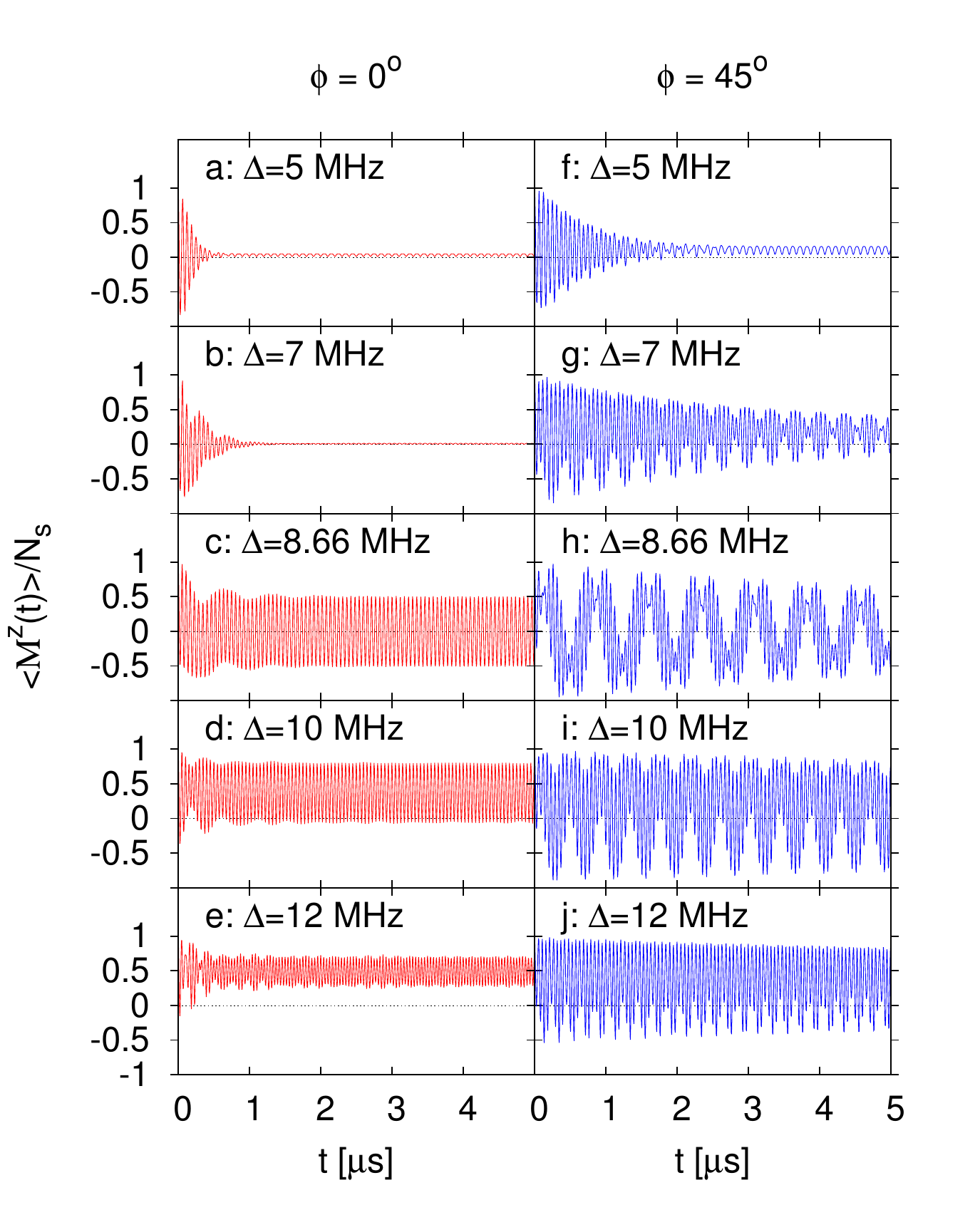}
\caption{(color online)
Simulation results for the total magnetization  obtained by solving the TDSE 
with Hamiltonian Eq.~(\ref{HB}) for a system of $N_{\mathrm{s}}=28$ interacting spins,
subject to a time-dependent magnetic field.
The spin-spin coupling strength $\lambda=5$, $h_d=15\,\hbox{MHz}$, and $h_i=0.12h_d=1.8\,\hbox{MHz}$.
\label{TDSEfig2a}
}
\end{figure}

\section{System of interacting magnetic moments}\label{section1b}

In the previous section, we studied effects of environments on the
spin dynamics and found decaying and sustained Rabi oscillations.
Regarding mechanisms for the magnetization decay, as a alternative
to the coupling to environment, we may also consider interactions
among the Gd$^{3+}$ moments and the homogeneity of the microwave field.
In this section, we study these two aspects by considering
a collection of spins, each one representing the magnetic moment of a Gd$^{3+}$ ion (and without additional spins representing a bath).
The model describes $N_\mathrm{S}$ spins that interact with the external magnetic fields and with each other.

The Hamiltonian of the whole system takes the form
\begin{eqnarray}
H(t)&=&\sum_{n=1}^{N_\mathrm{S}}
\big\{-\omega_0 S^z_n - 2\omega_d S^x_n\sin[2\pi(f_0+\Delta) t + \phi]
\nonumber \\
&&
-2\omega_i S^x_n\sin[2\pi(f_0-\Delta) t - \phi]
\big\}
\nonumber \\
&&
-\lambda\sum_{1=m<n}^{N_{\mathrm{S}}}\;\;\sum_{\alpha=x,y,z} C_{m,n,\alpha} S_{m}^\alpha S_{n}^\alpha
\;,
\label{SCB0b}
\end{eqnarray}
where the parameter $\lambda$ is used to set the overall strength of the coupling between the spins.
The first term in Eq.~(\ref{SCB0b}) describes the spins in the external time-dependent magnetic field.
The interaction between the spins is described by second term in Eq.~(\ref{SCB0b}).
The coefficients $C_{m,n,k}$'s are taken to be uniform random frequencies in the range $[-1\,\hbox{MHz},1\,\hbox{MHz}]$.
These randomly chosen $C_{m,n,k}$' s are assumed to mimic, in a simple manner, the essential features of the dipolar interactions between the Gd spins. 

As before, it is advantageous to eliminate the very fast motion associated with the Larmor
angular frequency $\omega_0$ by changing to the rotating frame.
The unitary transformation that accomplishes this is $U=\exp\left[i\omega_0(S^z_1+\ldots+S^z_{N_{\mathrm{S}}})\right]$.
Keeping only the secular terms, i.e. the terms that do not depend on $\omega_0$, the
interaction Hamiltonian does not change and Eq.~(\ref{SCB0b}) becomes
\begin{eqnarray}
H(t)&=&2\pi\sum_{n=1}^{N_{\mathrm{S}}}
\big\{
\Delta S^z_n
- h_d \left(S^x_n \sin\phi + S^y_n\cos\phi\right)
\nonumber \\
&&
\hbox to 1cm{}
+ h_i \left[S^x_n \sin(4\pi t\Delta+\phi) - S^y_n \cos(4\pi t \Delta+ \phi) \right]
\big\}
\nonumber \\
&&
-\lambda\sum_{1=m<n}^{N_{\mathrm{S}}}\;\;\sum_{\alpha=x,y,z} C_{m,n,\alpha} S_{m}^\alpha S_{n}^\alpha
\;.
\label{HB}
\end{eqnarray}
{\color{black}
As the $C$'s in Eq.~(\ref{HB}) are chosen uniformly random in the interval $[-1\,\hbox{MHz},+1\,\hbox{MHz}]$,
the average strength of the spin-spin interactions is
$\lambda\langle |C_{m,n,\alpha}|\rangle=\lambda/2 \,\hbox{MHz}$.
On the other hand, the characteristic energy scale of the spins is $2\pi h_d\approx94\,\hbox{MHz}$.
Therefore, for $\lambda=5$ (the value adopted in our simulations),
this strength is at least an order of magnitude smaller than the characteristic energy scale of the qubit.
In other words, the coupling between the spins may be considered to be weak.
}

{\color{black}
If we naively consider a uniform distribution of impurities over the host lattice with the concentration that we believe applies to the sample, we estimate the magnitude of lambda*C  (see Eq.(17)) to be approximately 0.3 MHz. This is a factor of 10 smaller than the average interaction strength (~2.5 MHz) used in our simulations. On the other hand, in our simulations, only a small (not more than 28 spins) cluster of spins suffice to create the "sustained Rabi oscillations". Assuming that average distance between the Gd$^{3+}$ ions in this cluster is only about two times smaller than for a uniform distribution Gd$^{3+}$ ions, changes the magnitude of magnitude of lambda*C to approximately 3 MHz. A slightly different, clustered distribution of impurities would suffice. Clearly, there is a lot of uncertainty in these estimates but at least they are not off by orders of magnitude.

Another aspect, not explicitly included in our simulation models is the following. As a spin system, Gd$^{3+}$ has $2S+1=8$ levels, two of which are effectively used in the ESR experiment. It is well-known, e.g. in the field of superconducting qubits~\cite{KOCH07,WILL17a}, that multilevel structures contribute to decoherence of the two-level system regarded as the qubit. The two-level systems of the bath are a very simple model that can mimic this mechanism. Of course, two-level systems can mimic ``almost anything'', as we also point out in the text. Also in this case, it is hard to put a reliable number on the interaction strength.
}

In the present case, to study the Rabi oscillations,
we solve the TDSE 
with Hamiltonian Eq.~(\ref{HB}) and take as the initial state
a the product state of all spins up (along the $z$-axis), that is
\begin{eqnarray}
|\Psi(t=0)\rangle&=& |{}\uparrow\rangle_1 \otimes \ldots \otimes|{}\uparrow\rangle_{N_{\mathrm{S}}}
\;.
\label{IS2}
\end{eqnarray}
Instead of the plotting the expectation value of the $z$-component of each spin,
we now plot $\langle M^z(t)\rangle=\sum_{n=1}^{N_{\mathrm{S}}} \langle \sigma^z_n(t)\rangle$, that is
the $z$-component of the total magnetization.

Figure~\ref{TDSEfig2a} shows the simulation data obtained by solving the TDSE 
for Hamiltonian Eq.~(\ref{HB}) with a system of $N_{\mathrm{S}}=28$ interacting spins.
Clearly, there is a significant qualitative difference between $\phi=0$ and $\phi=45^\circ$ data.
Furthermore, the latter does not compare well with the data shown in Fig.~\ref{FIGexp}(h--j).
For completeness, Figs.~\ref{A-SUP.TDSEfig2e} and~\ref{A-SUP.TDSEfig2f}
of the Supplemental Material~\cite{Note1} show the simulation
data for the time evolution of the energy of the model Eq.~(\ref{HB}).

\begin{figure}[!htp]
\centering
\includegraphics[width=\hsize]{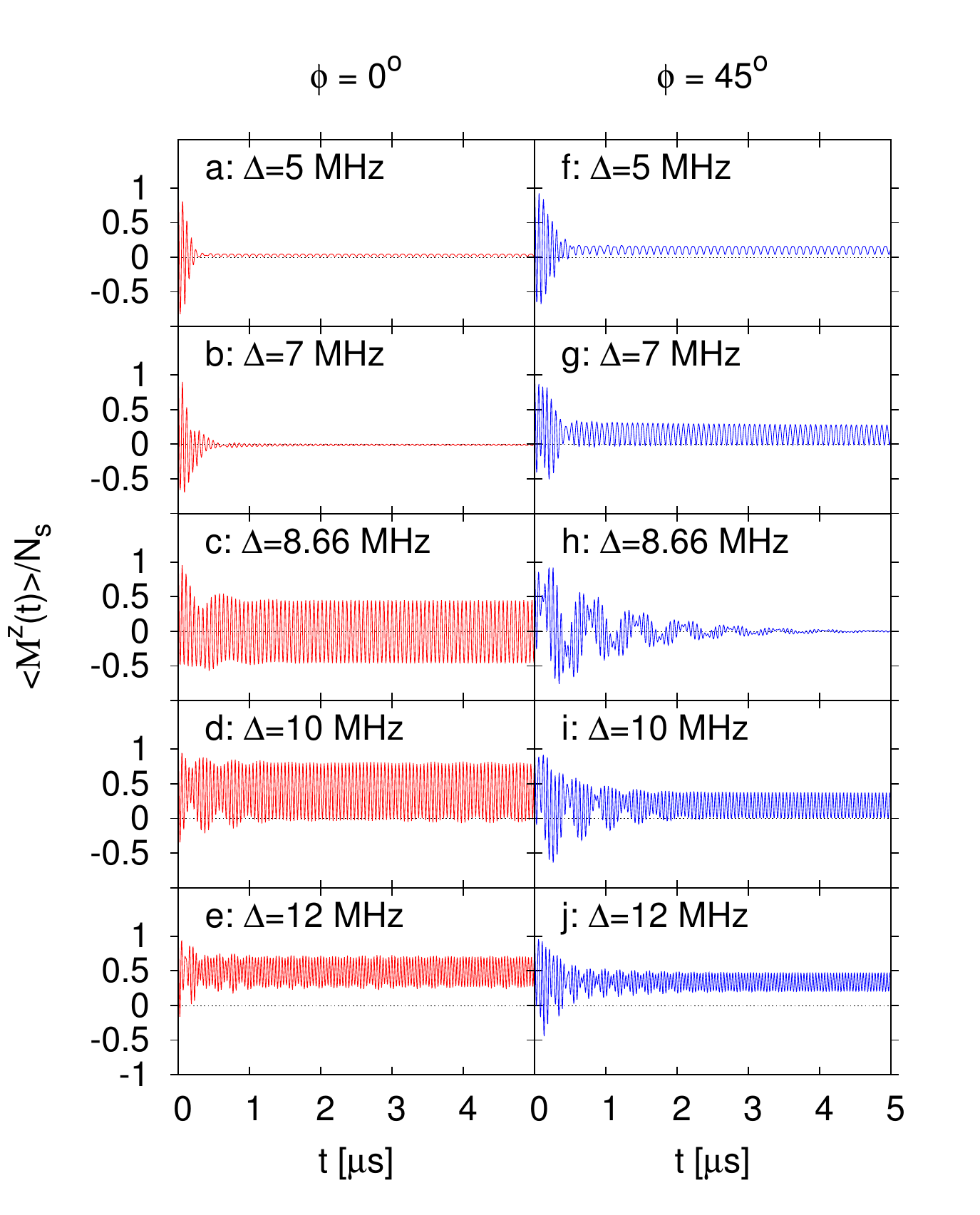}
\caption{(color online)
Simulation results obtained by solving the TDSE 
with Hamiltonian Eq.~(\ref{HC})
for a system of $N_{\mathrm{S}}=28$ interacting spins, spin-bath coupling $\lambda=5$ and
microwave field inhomogeneity $\delta=0.1$.
\label{INHOMfig3}
}
\end{figure}

\subsection{Inhomogeneity of the microwave fields}

The system of interacting spins,
a trivial modification allows us to account for the inhomogeneity of the microwave fields.
As we now show, the combination of spin-spin interactions and
the inhomogeneity of the microwave fields yields satisfactory results.
(for details see section~\ref{A-SUP.INHOM} of the Supplemental Material~\cite{Note1}).

Instead of Eq.~(\ref{HB}), we now take
\begin{eqnarray}
H(t)&=&2\pi\sum_{n=1}^{N_{\mathrm{S}}}\big\{
\Delta S^z_n
- h_{d,n} \left(S^x_n \sin\phi + S^y_n\cos\phi\right)
\nonumber \\
&&
\hbox to 1cm{}
+ h_{i,n} \left[S^x_n \sin(4\pi t\Delta+\phi) - S^y_n \cos(4\pi t \Delta+ \phi) \right]
\big\}
\nonumber \\
&&-\lambda\sum_{1=m<n}^{N_{\mathrm{S}}}\;\;\sum_{\alpha=x,y,z} C_{m,n,\alpha} S_{m}^\alpha S_{n}^\alpha
\;.
\label{HC}
\end{eqnarray}
In words, we place each spin $n$ in its own microwave fields $(h_{d,n},h_{i,n})$.
This modification has no impact on the computer resources it takes to solve the TDSE.
For the simulations reported in this paper,
the number of different $(h_{d,n},h_{i,n})$ pairs is equal to $N_{\mathrm{S}}=28$.

Figure~\ref{INHOMfig3} shows the simulation data obtained by solving the TDSE 
with Hamiltonian Eq.~(\ref{HC}) for a system of $N_{\mathrm{S}}=28$ interacting spins
in a inhomogeneous microwave field.
Although the inhomogeneity is weak ($\delta=0.1$), it has considerable impact on
the magnetization dynamics.

In the absence of spin-spin interactions ($\lambda=0$) and inhomogeneities,
all the spins perform the same motion, as discussed in section~(\ref{section1z}).
If $\lambda=0$, small differences in the microwave amplitudes $h_{d,n}$ and $h_{i,n}$, see Eq.~(\ref{HC}),
lead to quantitative but no qualitative differences in the motion of the spins.
However, if $\lambda\not=0$, the random interactions among the spins act against the coherent motion of the spins
and the Rabi oscillations decay on a short time scale, as illustrated in Fig.~\ref{INHOMfig3}(a,f),
except when the value of $\Delta$ satisfies the condition for the first Floquet resonance, see Fig.~\ref{INHOMfig3}(c,h).

\begin{figure}[!htp]
\centering
\includegraphics[width=\hsize]{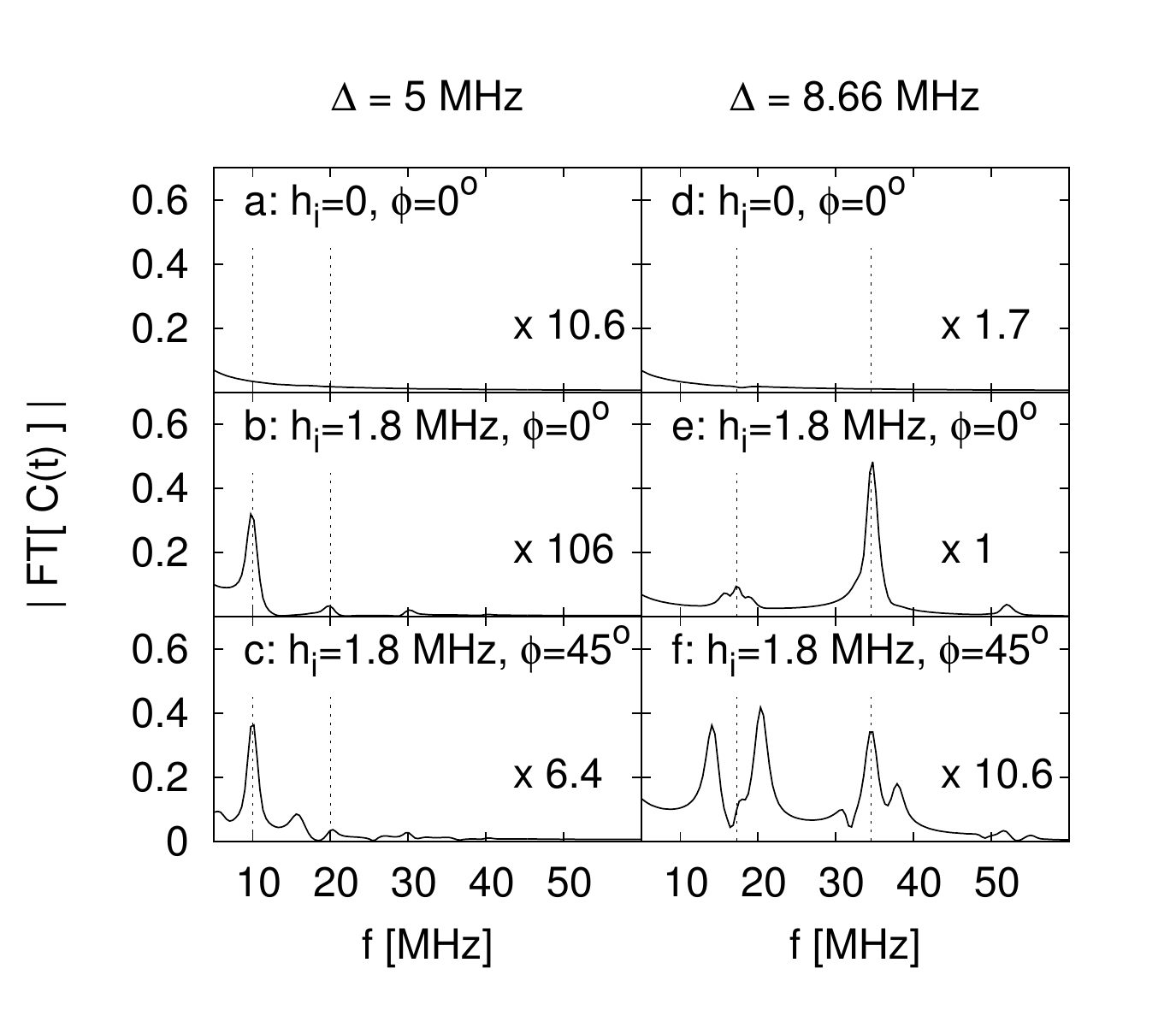}
\caption{
Simulation results of the Fourier transformed correlation Eq.~(\ref{SCB1a})
obtained by solving the TDSE 
for Hamiltonian Eq.~(\ref{HC}),
describing a system of $N_{\mathrm{S}}=28$ interacting spins,
The microwave field inhomogeneity $\delta=0.1$.
The positions of $2\Delta$ and $4\Delta$ are indicated by the two vertical dashed lines.
The plots (b), (c), (e) and (f) correspond to plots (a), (c), (f), and (h) in Fig.~\ref{INHOMfig3}, respectively.
At the Floquet resonance, the spins tend to develop a synchronized motion seen as a signal at $4\Delta$.
\label{INHOMfig4}
}
\end{figure}

In the case of the bath-I model, see section~\ref{section1a}, the spin and the bath-I spins establish some kind
of stationary state, see Fig.~\ref{TDSEfig1e}.
In the case of model Eq.~(\ref{HC}), the interactions among all the spins
may destroy the correlations between the individual spins as time proceeds.
In the present case, we take as a measure for these correlations
\begin{equation}
C(t)=\frac{2}{N_{\mathrm{S}}(N_{\mathrm{S}}-1)}\sum_{1=m<n}^{N_{\mathrm{S}}}\; \langle\Psi(t=0)| S_{m}^z(t) S_{n}^z(t)|\Psi(t=0) \rangle
\;,
\label{SCB1a}
\end{equation}
that is, the sum of all equal-time correlations of the $z$-components of all spins.
Here we do not present real-time data for $C(t)$ (see Fig.~(\ref{A-SUP.sumcor}))
but rather show its Fourier transform, see Fig.~\ref{INHOMfig4}.
As our main interest is in the long-time behavior of $C(t)$, we discard the transient regime
by considering real-time data for $1\,\mu\mathrm{s}\le t \le 5\,\mu\mathrm{s}$ only.

In the absence of the image drive $h_i=0$ (see Fig.~\ref{INHOMfig4}(a,d)),
the Fourier transformed $C(t)$ exhibits almost no structure and
has low intensity in comparison to the maximum intensities observed if the image drive is active
($h_i=1.8\,\mathrm{MHz}$, see Fig.~\ref{INHOMfig4}(b,c,e,f)).

If the image drive is present ($h_i=1.8\,\mathrm{MHz}$, $\Delta=5\,\mathrm{MHz}$
and $\phi=0$), Fig.~\ref{INHOMfig4}(b) suggests a correlated motion with a frequency of
$\approx 10\,\mathrm{MHz} \approx 2 \Delta$ (and some weak higher harmonics).
The spin system is outside the Floquet resonance regime and
therefore the only source of correlation is imposed by the image field drive,
similarly to the case of bath-I presented in the previous section.
Since the image drive runs at $2\Delta$, the correlation is driven at the same frequency.

This picture changes drastically if the image drive is present and $\Delta$
satisfies the condition for the first Floquet resonance $2\Delta=(h_d^2+\Delta)^{1/2}=F_R\approx17.3\,\mathrm{MHz}$.
Then Fig.~\ref{INHOMfig4}(e) suggest a correlated motion with a frequency of
$\approx35\,\mathrm{MHz} \approx 4 \Delta=2F_R$, that is a frequency of four
(not two as in the case $\Delta=5\,\mathrm{MHz}$) times the Rabi frequency.
Assuming that the $z$-component of each spin
is oscillating with some frequency, then multiplying two such components
(see $S_{m}^z(t) S_{n}^z(t)$ in Eq.~(\ref{SCB1a}))
gives rise to an oscillation with twice that frequency.
Thus, the clear signal at $\approx4 \Delta$ suggests
that $z$-component of each spin is oscillating with a frequency of $2\Delta$ imposed by the image drive and that there is some collective motion of these components.
In other words, at the Floquet resonance, the image drive induces
long-time correlations between the spins seen in the Fourier transform at a frequency of $2F_R$.
For $\phi=45^\circ$ the magnetization dynamics shows a much richer structure (see Fig.~\ref{INHOMfig3}(h))
which is also reflected in a more complicated spectrum, see Fig.~\ref{INHOMfig4}(g), but the signal
at $2F_R$ is also clearly visible.

The results the TDSE simulations for the model of interacting spins can be summarized as follows:
\begin{enumerate}
\item
Qualitatively, model Eq.~(\ref{HB}) does not reproduces the features of the Rabi oscillations shown in
Fig.~\ref{FIGexp}.
In particular, there is a clear qualitative difference
between Figs.~\ref{FIGexp}(h--j) and~\ref{TDSEfig2a}(h--j), that is if $\phi=45^\circ$.
Therefore, model Hamiltonian Eq.~(\ref{HB}) alone does not seem to describe well such oscillations.
\item
However, accounting for the inhomogeneity of the microwave field as in model Eq.~(\ref{HC}),
considerably improves the qualitative agreement with the experimental data (compare
Fig.~\ref{FIGexp} with Fig.~\ref{INHOMfig3}).
\item At the Floquet resonance, all interacting spins are developing a synchronised motion.
The effect disappears quickly when the resonance condition is not met.
\end{enumerate}

\section{Conclusion}\label{CONC}

Following its experimental implementation presented in Ref.~\cite{BERT20},
a spin qubit subject to two different microwave drives,
one that induces Rabi oscillations and a much weaker second one
aiming to preserve coherence was shown to extend the coherence time of a spin qubit.
In this paper, this protocol is analyzed theoretically by
numerically solving the time-dependent Schr\"odinger equation
of two different microscopic many-body systems.
The models considered are: (i) a spin qubit interacting with a collection of
two-level systems representing a bath (bath-I) and (ii) weakly interacting spin qubits
in which the spin qubits themselves act as a bath (bath-II).

We discuss the conditions for properly describing the measured Rabi
oscillations~\cite{BERT20} and analyze the microscopic internal dynamics
simulated for each type of bath. We find a small amount of saturation building
up in the spin-spin correlations of bath-I at long time scales
while the transitory regime shows signatures of the Floquet dynamics.
For the second model with bath-II, one clearly observe the build up of a
synchronised motion of the bath spins, sustained for very long times.
Both microscopic models are shown to be capable of reproducing
the main feature of the spin dynamics
observed in the ESR experiments on CaWO$_4$:Gd$^{3+}$.

The results of our simulation work suggest that in order to capture the essence of the
real-time dynamics of a spin as observed in the ESR experiments described above,
the effect of dynamics of this spin on the dynamics of the bath degrees of freedom has to be taken into account.
The dynamics of the bath degrees must be treated on the same footing as the dynamics of the spin itself.
{\color{black}
The build up of correlations between the spin and the bath spin result
in a synchronous motion of all the spins, having the effect of partial preservation of the phases.
}

We have demonstrated that with the computational resources that are available today,
it is possible to simulate the real-time dynamics of many-body quantum spin systems over a time span that is
experimentally relevant.
Our numerical simulations show how correlations are building up in the baths
when the Floquet protocol is implemented, leading to a preservation of the coherence of the qubit.
Both microscopic models are shown to be capable of reproducing
the main feature of the spin dynamics observed in the ESR experiments on CaWO$_4$:Gd$^{3+}$.

In view of the various physical mechanisms that may be at play,
it would be best to consider a model that simultaneously includes several of them.
Unfortunately, limitations imposed by computer resources
force us to consider the effects of different mechanisms separately.


\noindent
\medskip
\begin{acknowledgments}
The authors gratefully acknowledge the Gauss Centre
for Supercomputing e.V. (www.gauss-centre.eu) for funding
this project by providing computing time on the GCS
Supercomputer JUWELS at J\"ulich Supercomputing Centre (JSC).
Support from the project J\"ulich UNified
Infrastructure for Quantum computing (JUNIQ) that has received funding from the
German Federal Ministry of Education and Research (BMBF) and the Ministry of
Culture and Science of the State of North Rhine-Westphalia is gratefully acknowledged.
S.M. acknowledges support by the Elements Strategy Initiative Center for Magnetic Materials (ESICMM) (Grant No. 12016013) funded by the Ministry of Education, Culture, Sports, Science and Technology (MEXT) of Japan and also by the Grants-in-Aid for Scientific Research C (No.18K03444) from MEXT.
I.C. acknowledges partial support by the
National Science Foundation Cooperative Agreement No. DMR-1644779 and the State of Florida. Experimental data presented in  this  paper have  been  obtained with the help of the French research infrastructure IR INFRANALYTICS FR2054.
\end{acknowledgments}

\bibliography{../pumpingeffect,../floquet}

\end{document}